\documentclass[preprint,showpacs,preprintnumbers,amsmath,amssymb]{revtex4}

\usepackage{graphicx}
\usepackage{dcolumn}
\usepackage{bm}

\newcommand{\vv}[1]{\mbox{\boldmath{$#1$}}}

\begin{document}

\title{Spin polarization in a T-shape conductor \\
induced by strong Rashba spin-orbit coupling}

\author{Masayuki Yamamoto$^{1}$, Tomi Ohtsuki$^{2}$ and Bernhard Kramer$^{1}$}
\affiliation{
$^{1}$ I.Institut f\"{u}r Theoretische Physik, Universit\"{a}t Hamburg,
Jungiusstra{\ss}e 9, 20355 Hamburg, Germany \\
$^{2}$ Department of Physics, Sophia University,
Kioi-cho 7-1, Chiyoda-ku, Tokyo 102-8554, Japan}

\date{\today}

\begin{abstract}
We investigate numerically the spin polarization of the current in the
presence of Rashba spin-orbit interaction in a T-shaped conductor
proposed by A.A. Kiselev and K.W. Kim (Appl. Phys. Lett. {\bf 78} 775
(2001)).
The recursive Green function method is used to calculate the three terminal
spin dependent transmission probabilities.
We focus on single-channel transport and show that the spin polarization
becomes nearly 100\,\% with a conductance close to $e^{2}/h$ for sufficiently
strong spin-orbit coupling.
This is interpreted by the fact that
electrons with opposite spin states are deflected into an opposite terminal
by the spin dependent Lorentz force.
The influence of the disorder on the predicted effect is also discussed.
Cases for multi-channel transport are studied in connection with experiments.
\end{abstract}

\maketitle

\section{Introduction}

Recently, spin-dependent electronic transport is attracting considerable
attention because of possible applications to spintronics \cite{Wolf,Zutic}.
Many of the proposals for two dimensional (2D) spintronic devices are based on
the presence of spin-orbit coupling in the 2D electron system (2DES)
semiconductor heterostructure.
There are two types of spin-orbit coupling terms in such systems.
One is the so-called Dresselhaus term which originates from the inversion
asymmetry of the zinc-blende structure \cite{Dresselhaus}.
The other is described by the Rashba Hamiltonian, 
\begin{equation}
{\cal H}_{\mathrm R} = 
\frac{\alpha}{\hbar} (\sigma_{x}p_{y} - \sigma_{y}p_{x})
\label{eq:Rashba}
\end{equation}
where $\alpha$ denotes the strength of spin-orbit coupling, $\sigma_{i}$ and
$p_{i}$ ($i=x,y$) are the Pauli matrices and the components of the momentum,
respectively.
The Rashba mechanism is due to the effective electric field originating from the
asymmetry of the potential confining the 2DES \cite{Rashba60,Rashba84}.

It is well known that the Rashba term dominates in narrow-gap semiconductors
while the Dresselhaus term is dominant in wide-gap systems
\cite{Lommer}.
Since the strength of the Rashba term can be controlled via external gates
\cite{Nitta,Engels}, 2DESs with Rashba spin-orbit interaction have become most
promising for spintronic applications.

In order to realize such devices, one needs spin polarized electrons in the
semiconductor inversion layer.
Most straightforwardly, one could generate spin polarized electrons by
attaching ferromagnetic metallic contacts to the 2DES and by injecting a current
\cite{Datta}.
However, it has been found that in practice the efficiency of the spin
injection from a ferromagnet into a semiconductor is very poor because
of the conductivity mismatch \cite{Schmidt}.
Thus, alternative methods have to be invented. 
Since strong spin-orbit scattering can lead to spatially inhomogeneous 
spin polarization \cite{Landaulifshits} the generation of spin polarized 
electrons via spin-orbit coupling is in principle possible.

In this paper we consider a conductor with a T-shape structure with Rashba
spin-orbit coupling as originally proposed by Kiselev and Kim \cite{Kiselev}.
For relatively small strength of the spin-orbit coupling, these authors have
obtained high spin polarization.
However, the corresponding conductance has been found to be very small.
This problem of the small conductance has been eventually overcome
by considering a ring-shape electron resonator
\cite{Kiselev-ring}.
It has also been reported that spin accumulation occurs for considerably
strong spin-orbit coupling in a quasi-1D wire \cite{Governale}.

In both of the above cases, electron transport in the lowest subband
(single-channel transport) has mainly been considered.
Due to the self-duality of scattering matrix \cite{Beenakker} for the system
with spin-orbit interaction, no spin polarization of the current can be
obtained for single-channel transport in two-terminal devices.
Therefore, one has to consider at least a conductor with three terminals.
The single-channel limit seems to be ideal because one can completely suppress
the effect of the D'yakonov-Perel' relaxation, which is the major spin
relaxation mechanism in such systems
\cite{Malshukov,Kiselev-wire,Nikolic-DP,Pramanik}.
%

%
We will show in this paper that the amplitude of the spin polarization becomes 
almost
perfect with only very little loss of the conductance if the spin-orbit
interaction is sufficiently strong. We argue that the predicted effect should
be experimentally accessible in InAs.

In the next section, we describe the model system to be investigated
numerically. We use the Ando tight-binding Hamiltonian with Rashba coupling
in the off-diagonal matrix elements \cite{Ando89}.
In Sec. \ref{sec:result}, the results for the dependence of the polarization
on the energy and the strength of spin-orbit coupling are presented.
The amplitude of spin polarization is shown to depend on the ratio between the
$\pi$-phase spin precession length and the width of the quantum wires of
the T-shape conductor.
In Sec. \ref{sec:discuss}, 
we discuss the origin of the polarization by investigating
the spin states of the wave function.
We show that the propagating electrons are deflected at the junction by 
^^ ^^ Lorentz force" due to the spin-orbit induced effective magnetic field 
proportional to the $z$-component of the spin state.
The effects of disorder and other channels are also investigated.
The final section is devoted to the summary of this paper and the
discussion for experimental realization.

\section{The Model}

\begin{figure}[ht]
\includegraphics[scale=0.55]{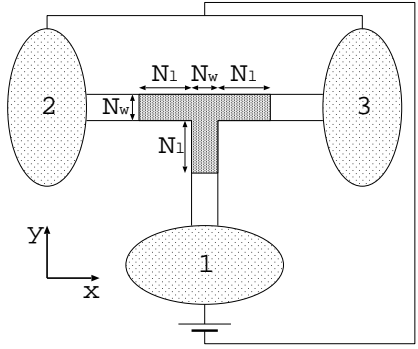}
\caption{
  Schematic view of the T-shaped conductor. Current injected from reservoir 1 
can
  go to reservoirs 2 or 3. Shaded: regions with non-vanishing
  spin-orbit coupling; parameters: $N_{w}=10a$ and $N_{l}=20a$ with $a$ lattice
  spacing of tight-binding model.
\label{fig:Tshape}}
\end{figure}

We consider the T-shape conductor shown in 
Fig.\ref{fig:Tshape} in the presence of Rashba spin-orbit coupling.
The sample region is connected to three electron reservoirs by ideal leads.
Electron current is injected into the sample from the reservoir 1 and 
goes to reservoirs 2 or 3.
At small voltages, the currents $I_{21}$ and $I_{31}$ from reservoir 1 to
reservoirs 2 and 3, respectively, are proportional to the conductances
$G^{21}$ and $G^{31}$.

The Rashba spin-orbit coupling can be described in the tight-binding
language by the Ando Hamiltonian \cite{Ando89},
\begin{equation}
{\cal H} = \sum_{i,\sigma}W_{i} c_{i\sigma}^{\dagger}c_{i\sigma}
- \sum_{\langle i,j \rangle ,\sigma,\sigma'} V_{i\sigma,j\sigma'}
c_{i\sigma}^{\dagger}c_{j\sigma'} \,,
\label{eq:Ando}
\end{equation}
where
\begin{equation}
V_{i,i+\hat{x}} = V_0
\left( \begin{array}{cc}
\cos \theta  & -\sin \theta \\
\sin \theta  &  \cos \theta \\
\end{array} \right) \,,
\label{eq:x-hopping}
\end{equation}
and
\begin{equation}
V_{i,i+\hat{y}} = V_0
\left( \begin{array}{cc}
\cos \theta   & i \sin \theta \\
i \sin \theta &   \cos \theta \\
\end{array} \right).
\label{eq:y-hopping}
\end{equation}
Here, $c_{i\sigma}^{\dagger} (c_{i\sigma})$ denotes the creation (annihilation)
operator of an electron on site $i$ with spin $\sigma$, $W_{i}$ the random
potential on the site $i$ distributed uniformly in $[-W/2,W/2]$, and
$V_{i,i+\hat{x}}(V_{i,i+\hat{y}})$ the hopping matrix elements in $x$-($y$-)
directions.
The hopping is restricted to nearest neighbours.
The hopping energy $V_0 = {\hbar^2}/{2m^* a^2}$, where $m^*$ is the effective
electron mass and $a$ the tight-binding lattice spacing, is taken as the unit
of the energy.
The parameter $\theta$ represents the strength of the spin-orbit coupling and
is related to Rashba coupling $\alpha$ by
\begin{equation}
\alpha \simeq 2\theta V_{0}a 
\hspace{1cm} (\hspace{0.1cm} {\rm for} \hspace{0.2cm} \theta \ll 1).
\end{equation}

The conductance between reservoirs $J$ to $I$ \cite{Landauer}
and the corresponding spin polarization are defined by
\begin{equation}
 G^{IJ} = G_{0} {\rm Tr}\, t_{IJ}^{\dagger} t_{IJ} ,
\label{cond-eq}
\end{equation}
and
\begin{equation}
 P_{k}^{IJ} = 
\frac{{\rm Tr}t_{IJ}^{\dagger} \sigma_{k} t_{IJ}}
{{\rm Tr} t_{IJ}^{\dagger} t_{IJ}}
\hspace{0.5cm} (k = x,y,z) \,,
\end{equation}
with the conductance quantum $G_{0} \equiv {e^{2}}/{h}$, $t_{IJ}$ denoting the
transmission matrix from reservoirs $J$ to $I$.
Below, we will focus on the transport between reservoirs 1 and 2. Transport
between reservoirs 1 and 3 can trivially be deduced via current conservation
and the symmetry of the system.
We calculate the amplitude of the total spin polarization defined by
\begin{equation}
|\bm{P}|=({P_{x}^{2}+P_{y}^{2}+P_{z}^{2}})^{1/2},
\end{equation}
instead of considering only the $z$-component.

As described in detail in Appendix \ref{sec:algo},
we can obtain the transmission coefficient $t_{\mu \nu}^{IJ}$
for the incident channel $\nu$ with velocity $v_{\nu}$ in the probe $J$
and out-going channel $\mu$ with velocity $v_{\mu}$ in the probe $I$
as \cite{Ando91}
\begin{equation}
t_{\mu \nu}^{IJ} = \left(\frac{v_{\mu}}{v_{\nu}}\right)^{1/2}
[-(U^I)^{-1} \hat{G}^{IJ} U^J \{(\Lambda^J)^{-1} - \Lambda^J \}]_{\mu \nu}
\end{equation}
where $\hat{G}^{IJ}$ is Green function
corresponding to the transport from the probe $J$ to $I$.
$\Lambda^J$ contains only diagonal elements,
$\Lambda^J(i,j)=\lambda_i^J \cdot \delta_{ij}$,
where $\lambda_i^J$ is the $i$-th eigenvalue 
of the transfer matrix for an ideal region in the probe $J$
and $U^{I(J)}$ consists of the set of eigenfunctions 
for \{$\lambda_i^{I(J)}$\}.

\section{Results \label{sec:result}}

Using the recursive Green function method,
we calculate the amplitude of the spin polarization.
For the system size we assume $N_{w}=10a$ and $N_{l}=20a$.
We first consider an clean system ($W=0$).

\begin{figure}[ht]
\includegraphics[scale=0.35]{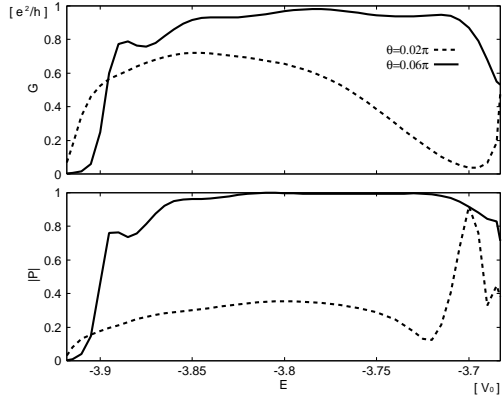}
\caption{
  Conductance $G$ and spin polarization $|P|$ as functions of the Fermi
  energy $E$ in the region of single-channel transport. Almost 100\%
  polarization is obtained together with $G\approx G_{0}$ for stronger
  spin-orbit coupling ($\theta = 0.06\pi$). For $\theta = 0.02\pi$ high
  polarization is obtained only at energies just before the second channel
  opens.
\label{fig:P_vs_E}}
\end{figure}

Figure \ref{fig:P_vs_E} shows the dependence of 
the conductance and the spin polarization on the Fermi energy $E$
for weak and strong spin-orbit couplings.
We consider the energy region for single-channel transport.
For weaker spin-orbit coupling ($\theta = 0.02\pi$), high spin polarization is
obtained for energies just before the second channel opens ($E \simeq
-3.68V_0$). The corresponding conductance is small as compared to $G_{0}$
\cite{Kiselev}.
Almost perfect polarization is obtained together with a
conductance close to $G_{0}$ for stronger spin-orbit coupling ($\theta =
0.06\pi$).
Here, the polarization is almost insensitive to the energy except near the
band edge.

\begin{figure}[ht]
\includegraphics[scale=0.35]{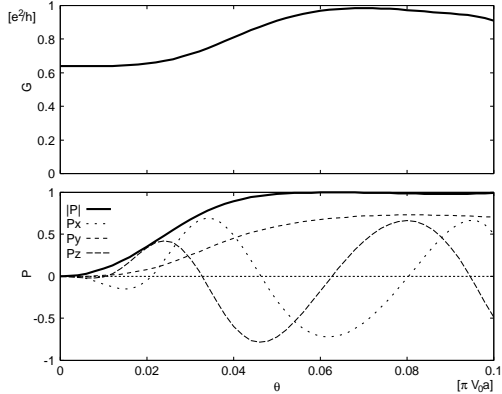}
\caption{
  Conductance $G$ and spin polarization $P_{k}$ in the directions of
  $k=x,y,z$ as functions of the strength of spin-orbit coupling $\theta$ for
  $E=-3.8V_{0}$; $P_{y}$ increases monotonically with increasing
  $\theta$ while $P_{x}$ and $P_{z}$ oscillate.
\label{fig:P_vs_SO}}
\end{figure}

Figure \ref{fig:P_vs_SO} shows the dependence of the conductance and the spin
polarization on the strength of spin-orbit coupling at energy
$E=-3.8V_0$.
With the increase of the strength of spin-orbit coupling, $P_{y}$ also increases
monotonically while $P_{x}$ and $P_{z}$ oscillate.
We also note that the conductance increases together with the amplitude of the
polarization.
Due to the symmetry of the T-shaped conductor, the current into reservoir 3
has the same polarization in the direction of $y$ but opposite polarizations
in $x$- and $z$-directions \cite{Kiselev}.

\begin{figure}[ht]
\includegraphics[scale=0.4]{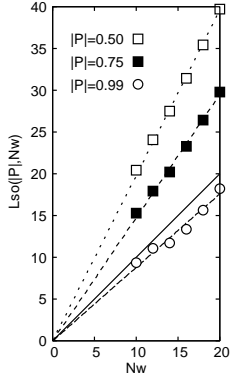}
\caption{
The spin precession lengths $L_{\rm so}$ for $|P|=0.5, 0.75, 0.99$ as
functions of the width of the wires $N_w$.
Energy is set to be the middle of lowest and first excited subband,
e.g., $E=-3.8 V_{0}$ for $N_w=10$.
The polarization becomes almost perfect 
if the spin precession length becomes shorter than the wire width, 
$L_{\rm so}(|P|,N_w)={\pi a}/{2\theta(|P|,N_w)}< N_w$.
Solid line: $L_{\rm so}(|P|,N_w)= N_{w}$ corresponding to $|P| \simeq 0.97$.
\label{fig:Lso_vs_Nw}}
\end{figure}

\begin{figure}[ht]
\includegraphics[scale=0.35]{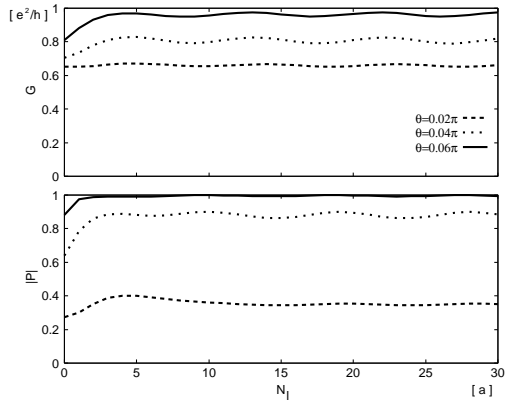}
\caption{
Conductance $G$ and spin polarization $|P|$ 
as functions of the wire length $N_l$
for several strengths of the spin-orbit coupling at $E=-3.8V_0$.
The polarizations are insensitive to the length of leads
except for very short length.
\label{fig:P_vs_Nl}}
\end{figure}

What are the  conditions for achieving almost perfect polarization?
We define the $\pi$-phase spin precession length 
$L_{\rm so}(|P|,N_w) = {\pi a}/{2\theta(|P|,N_w)}$ 
where $\theta(|P|,N_w)$ is the spin-orbit coupling strength
giving rise to the polarization $|P|$ for the width $N_w$.
From the plot $L_{\rm so}(|P|,N_w)$ as a
function of the width of the system $N_w$ (Fig.~\ref{fig:Lso_vs_Nw})
one concludes that $L_{\rm so}(|P|,N_w)$ is almost linear in $N_w$ and high
polarization, $|P| > 0.99$, is achieved for $L_{\rm so}(|P|,N_w) < N_w$.
On the other hand, no dependence on the length of the leads ($N_l$) is
observed as shown in Fig.\ref{fig:P_vs_Nl}.

\section{Discussion \label{sec:discuss}}
It is well known that the spin-orbit interaction is the origin 
of the anomalous Hall effect in ferromagnetic materials
\cite{Karplus}.
This is because the spin-orbit coupling affects polarized conduction
electrons as a spin-dependent effective magnetic field.
%
%
In this section, we investigate the spin wave function
and show that propagating electrons are deflected at the junction by 
the Lorentz force due to the spin-orbit induced effective magnetic field 
proportional to the $z$-component of the spin \cite{Karplus,li05,Nikolic}.
We also investigate the effect of disorder 
and the influence of other transport channels.

\subsection{Spin-orbit induced effective magnetic field}
A convenient way of deriving the effective magnetic field is to estimate the 
flux
per plaquette from the Aharonov--Bohm phase.  Let an electron initially be at 
$(i,j)$.
When it hops from $(i,j)$ to $(i+\hat{x},j)$ to $(i+\hat{x},j+\hat{y})$ 
according to
Eqs.~(\ref{eq:x-hopping}) and (\ref{eq:y-hopping}),
the SU(2) phase 
$\exp(-{\mathrm{i}}\theta\sigma_x)\exp({\mathrm{i}}\theta\sigma_y)$
is acquired.  On the other hand, when it hops from $(i,j)$ to $(i,j+\hat{y})$ to
$(i+\hat{x},j+\hat{y})$, the phase it obtains is
 $\exp({\mathrm{i}}\theta\sigma_y)\exp(-{\mathrm{i}}\theta\sigma_x)$.
Therefore the interference between the two path along the plaquette becomes
\begin{equation}
\left(\exp({\mathrm{i}}\theta\sigma_y)\exp(-{\mathrm{i}}\theta\sigma_x)\right)^\dagger
\exp(-{\mathrm{i}}\theta\sigma_x)\exp({\mathrm{i}}\theta\sigma_y)
=
\exp({\mathrm{i}}\theta\sigma_x)\exp(-{\mathrm{i}}\theta\sigma_y)
\exp(-{\mathrm{i}}\theta\sigma_x)\exp({\mathrm{i}}\theta\sigma_y)\,.
\end{equation}
This can be evaluated
on the basis of Campbell-Hausdorff formula,
\begin{equation}
\exp(\theta X)\exp(\theta Y)=\exp\left(\theta (X+Y)+
\frac{\theta^2}{2}[X,Y]+O(\theta^3)\right)\,,
\end{equation}
to find
\begin{eqnarray}
& &\exp({\mathrm{i}}\theta\sigma_x)\exp(-{\mathrm{i}}\theta\sigma_y)
\exp(-{\mathrm{i}}\theta\sigma_x)\exp({\mathrm{i}}\theta\sigma_y)\\\nonumber
&\approx&\exp({\mathrm{i}}(\sigma_x-\sigma_y)+{\mathrm{i}}\theta^2\sigma_z)
\exp(-{\mathrm{i}}\theta\sigma_x)\exp({\mathrm{i}}\theta\sigma_y)\\\nonumber
&\approx&\exp({\mathrm{i}}(\sigma_x-\sigma_y-\sigma_x)+2{\mathrm{i}}\theta^2\sigma_z)
\exp({\mathrm{i}}\theta\sigma_y)\\\nonumber
&\approx&\exp(2{\mathrm{i}}\theta^2\sigma_z)\,,
\end{eqnarray}
where we have dropped terms higher than $\theta^2$.
Using the relation,
\begin{equation}
2\pi B a^2/(h/e)=2\theta^2\sigma_z\,,
\end{equation}
we have \cite{li05,Nikolic}
\begin{equation}
B=\frac{\hbar}{e}\frac{2\theta^2}{a^2}\sigma_z\,.
\label{eq:Beff}
\end{equation}

\begin{figure}[ht]
\includegraphics[scale=0.8]{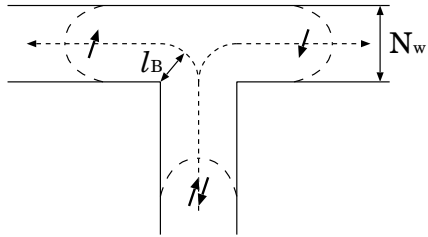}
\caption{
Schematic view of the electron trajectory.
A pair of the electrons with anti-parallel spins are almost completely
separated at the junction when the cyclotron diameter becomes shorter 
than the wire width ($2l_{\bar{B}}(\theta) < N_w$).
\label{fig:l_B}}
\end{figure}

We now discuss the condition for the high polarization
in terms of this spin dependent Lorentz force.
We use Eq.(\ref{eq:Beff}) to estimate the strength of magnetic field
\begin{equation}
\bar{B}(\theta)=\frac{4}{\pi} \frac{\hbar}{e}
\left( \frac{\theta}{a} \right)^2.
\end{equation}
where we have averaged $\sigma_z$
to be $2/\pi$, since the variation of the expectation value of $\sigma_z$
is described by the $\cos$--function.
Since the kinetic energy is comparable to the confinement energy
in single-channel transport,
the velocity of an injected electron can be assumed to be
\begin{equation}
v \simeq \frac{\hbar}{m^*} \frac{\pi}{N_w}.
\end{equation}
Then the corresponding cyclotron diameter is given by
\begin{equation}
2l_{\bar{B}} (\theta) = \frac{2m^* v}{e \bar{B}(\theta)}
= \frac{2L_{\rm so}^2}{N_w}.
\end{equation}
The cyclotron diameter becomes shorter than the wire width
($2l_{\bar{B}}(\theta) < N_w$)
if the spin precession length becomes shorter than the wire width
($L_{\rm so} < N_w/\sqrt{2}$).
As a result, electrons with opposite $z$--component spin
are almost completely separated at the junction and nearly perfect spin
polarization is obtained (Fig.\ref{fig:l_B}).
This situation is similar to the mesoscopic cross junction in magnetic
fields \cite{beenakker89}.

\subsection{The influence of disorder}

\begin{figure}[ht]
\includegraphics[scale=0.35]{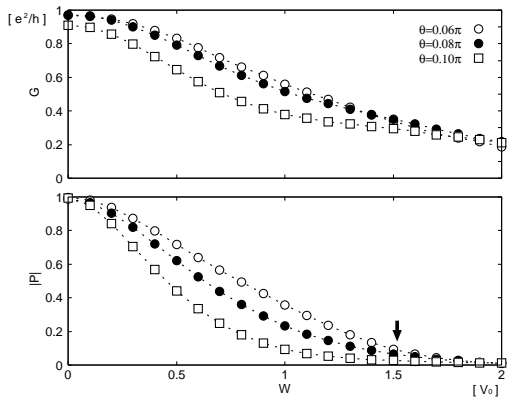}
\caption{
  Conductance $G$ and spin polarization $|P|$ as functions of the strength of
  disorder $W$ for several spin-orbit couplings at $E = -3.8V_{0}$.
  Ensemble average has been taken over $10^4$ samples. For stronger
  spin-orbit coupling, the polarization becomes more sensitive to disorder.
  Arrow: crossover between ballistic and diffusive regimes [cf.
  Eq.(\ref{eq:mfp})].
\label{fig:P_vs_W}}
\end{figure}

We now consider briefly the effect of disorder on the spin polarization
(Fig.\ref{fig:P_vs_W}).
An ensemble average is performed over $10^{4}$ samples.
The suppression of the polarization by disorder becomes more prominent
as the spin-orbit coupling becomes stronger.
The mean free path of a 2DES in the tight-binding  model is described by
\cite{Ando91}
\begin{equation}
L_{m} = 48 a V_{0}^{3/2} \frac{\sqrt{E+4V_{0}}}{W^2}.
\label{eq:mfp}
\end{equation}
One can use this estimate to distinguish the ballistic regime from the
diffusive one.  For the present system, we obtain
$W \simeq 1.53V_{0}$ for $L_{m}=50a$ (indicated by an arrow in 
Fig.~\ref{fig:P_vs_W}).
As seen in the figure, the sample size must be smaller than the mean free path
in order to obtain high spin polarization.

\subsection{The influence of other channels \label{sec:multi-ch}}

\begin{figure}[ht]
\includegraphics[scale=0.35]{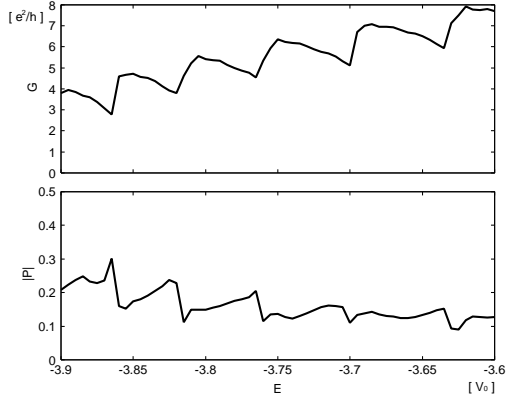}
\caption{
Conductance $G$ and spin polarization $|P|$ 
as functions of the Fermi energy $E$
in the region of multi-channel transport.
The system size is set to be $N_w = 50a$ and $N_l = 50a$, and
the strength of spin-orbit coupling $\theta = 0.01\pi$.
The number of channel increases to values ranging from 5 to 10
in this energy region.
The polarization stays higher than 10\%.
\label{fig:nw50nl50}}
\end{figure}

Finally let us consider the system
whose size is $N_w = 50a$ and $N_l = 50a$.
Figure \ref{fig:nw50nl50} shows the dependence of the conductance and
the spin polarization on the Fermi energy $E$ for $\theta = 0.01\pi$.
In this energy region
where the number of channels increases to values ranging from 5 to 10,
several channels contribute to transport.
While the spin polarization is reduced by channel mixing,
it still stays higher than 10\%.

\section{Summary}

In this paper we have investigated numerically spin polarized linear transport
in a T-shaped conductor with Rashba spin-orbit coupling.
We have considered single-channel transport and found that the spin
polarization becomes almost perfect with a conductance close to $G_{0}$ for
stronger spin-orbit coupling.
Since the spin-orbit coupling can be regarded as
the spin-dependent effective magnetic field \cite{Karplus},
the propagating electrons with certain initial spin states 
are deflected into one of the two output terminals by Lorentz force
while those with opposite spin states are deflected into the other.
The ratio between the cyclotron diameter in an effective field 
and the wire width
depends on that between the $\pi$-phase spin precession length 
and the wire width.
If the precession length becomes shorter than the wire width, 
the cyclotron diameter becomes shorter than the wire width 
so that electrons with antiparallel spins 
are almost completely separated at the junction and nearly 100\%
polarization can be achieved.

With respect to the intrinsic spin Hall effect \cite{Murakami},
we should note the following point.
%
The spin-orbit induced effective magnetic field can cause
the anomalous Hall effect in ferromagnets, where the spin direcion of 
conduction electrons is maintained by the interaction between 
localized spin states.
However, it can not simply separate injected electrons with
up and down spins parallel to the $z$-axis in the investigated system
since the spin states of propagating electrons 
are always changing due to the spin precession 
\cite{Nikolic}.

We have also investigated the effect of disorder on the polarization.
Since the polarization becomes more sensitive to the disorder when the
spin-orbit interaction is stronger, one needs to prepare the clean samples so
that the system belongs to the ballistic regime to realize the perfect
polarization of the current.

In order to obtain information 
whether or not the predicted effects are observable in experiment,
let us consider the parameters required for one of 
the favorable materials, InAs,
with an effective mass $m^* =0.039m_0$ ($m_0$ is the free electron mass)
and Rashba coupling $\alpha = 23.8 \times 10^{-12}\,$eVm \cite{Hu}.
Let us assume that the width of the conduction band is
$\Delta=$1\,eV.
This gives for $V_{0}=\Delta/2Z=125\,$meV ($Z=4$ for square lattice),
and for the tight-binding lattice parameter $a=\hbar/(2m^{*}V_{0})^{1/2}\simeq 
2.8\,$nm.
Using the above numerical value of $\alpha$ 
one obtains $\theta=\alpha/2V_{0}a \simeq 0.01\pi$.
This would reduce the polarization to about 10\%,
still a reasonable value for being observable in experiment.
The crucial point, however, is the condition that the transport 
has to be in the single-channel regime.
The wire width should be about $20\,$nm 
for single-channel transport
when Fermi energy is of the order of $10\,$meV.
In principle it is possible to fabricate such a narrow wire,
but it makes the mean free path shorter
and the effect of disorder may become critical.
On the other hand, the wire width becomes $140\,$nm 
for the system with $N_w = 50a$ (Sec.\ref{sec:multi-ch}).
This width of quantum wire can be easily fabricated than
that for single-channel transport, which indicates that we can
observe the spin polarization experimentally.

\begin{acknowledgments}
We thank J. Nitta, J. Ohe, K. Dittmer, K. Slevin and S. Murakami
for fruitful discussions.
This work was supported by SFB 508 ^^ ^^ Quantenmaterialien" 
of the Deutsche Forschungsgemeinschaft,
the Marie-Curie Network MCRTN-CT2003-504574 of the EU,
and the Grant-in-Aid 14540363
from the Ministry of Education, Culture,
Sports, Science and Technology.
\end{acknowledgments}

\appendix
\section{Recursive Green function method for multi-terminal geometry
\label{sec:algo}}

\begin{figure}[ht]
\includegraphics[scale=0.6]{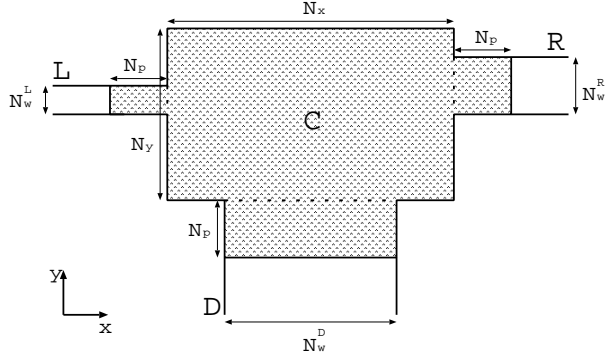}
\caption{\label{fig:multi-term}
Schematic draw of three terminal geometry.
The shaded area represents a sample region
where spin-orbit interaction is present.
This sample region is connected to three electron reservoirs
by ideal probes.
}
\end{figure}

For numerical calculations,
we apply the recursive Green function method \cite{Ando91}
to the case of multi-terminal geometry.
Let us consider three terminal geometry described in Fig.\ref{fig:multi-term}.
The central sample region ($C$) is attached 
to three probes ($D,R$ and $L$).
Each probe consists of the infinite ideal region
and the sample one
whose size is $N_p \times N_w^I$ ($I=D,R$ and $L$).

\begin{figure}[ht]
\includegraphics[scale=0.6]{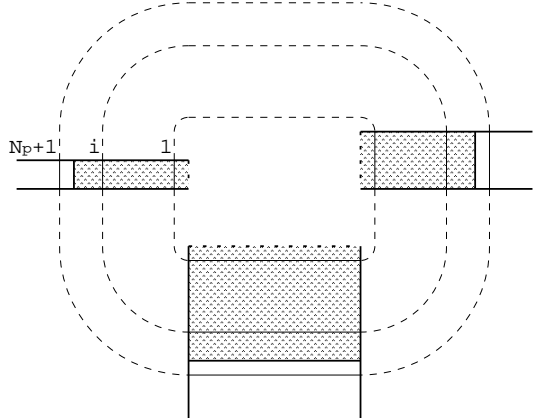}
\caption{\label{fig:slice}
Graphical interpretation for the Hamiltonian ${\cal H}_i$.
The $i$-th slices of three probes are gathered.}
\end{figure}

The full Hamiltonian can be written down as 
\begin{eqnarray}
\widetilde{\cal H}
=
\left(
\begin{array}{ccccccc}
\widetilde{\cal H}_{N_p +1} & -V_{N_p +1,N_p} & 0          &\cdots&0&0&0\\
-V_{N_p +1,N_p}^{\dagger} & {\cal H}_{N_p} & -V_{N_p,N_p -1} &\cdots&0&0&0\\
0 & -V_{N_p,N_p -1}^{\dagger} & {\cal H}_{N_p -1} &\cdots&0&0&0\\
\vdots & \vdots & \vdots & \ddots & \vdots & \vdots & \vdots\\
0&0&0&\cdots& {\cal H}_{2} & -V_{2,1} & 0 \\
0&0&0&\cdots& -V_{2,1}^{\dagger} & {\cal H}_{1} & -V'_{1,C} \\ 
0&0&0&\cdots& 0 & -{V'}_{1,C}^{\dagger} & {\cal H}^{C} \\
\end{array}
\right).
\label{eq:effectH}
\end{eqnarray}
${\cal H}^{C}$ is the Hamiltonian for the central sample region 
and can be described by
\begin{eqnarray}
{\cal H}^C
=
\left(
\begin{array}{ccccc}
{\cal H}_1^C & -V_{12}^C & 0         &\cdots&0\\
-V_{21}^C & {\cal H}_2^C & -V_{23}^C &\cdots&0\\
0 & -V_{32}^C & {\cal H}_3^C         &\cdots&0\\
\vdots & \vdots & \vdots & \ddots & \vdots\\
0 & 0 & 0 & \cdots & {\cal H}_{N_x}^C
\end{array}
\right)\,,
\label{}
\end{eqnarray}
where ${\cal H}_i^C$ denotes the Hamiltonian 
for the $i$-th slice along the $y$-direction and
$V_{ij}^C$ the hopping term between the slice $i$ and $j$.
The hopping is restricted to nearest neighbours.
$V'_{1,C}$ is the hopping term between 
the central sample region and its neighbouring slices of three probes.
${\cal H}_i$ is the Hamiltonian for the set of $i$-th slices of three probes
and can be described by 
\begin{eqnarray}
{\cal H}_{i}=\left(
\begin{array}{ccc}
 {\cal H}_{i}^{D} &0&0\\
0& {\cal H}_{i}^{R} &0\\
0&0& {\cal H}_{i}^{L} \\
\end{array}
\right)\,,
\label{eq:Hprobe}
\end{eqnarray}
where ${\cal H}_{i}^{I}$ ($I=D,R$ and $L$) is the Hamiltonian 
for the $i$-th slice of the probe $I$ (see Fig.\ref{fig:slice}).
$V_{ij}$ is the hopping term between 
the set of $i$-th slices and $j$-th ones.

Let us define some variables by following Ref.\cite{Ando91} as
\begin{eqnarray}
\Lambda^I=\left(
\begin{array}{ccc}
{\lambda}_{1}^{I} &&\\
& \ddots &\\
&& {\lambda}_{N_w^I}^{I}\\
\end{array}
\right) \,,
\end{eqnarray}
and 
\begin{equation}
U^I = \left( \vv{u}_1^I  \cdots \vv{u}_{N_w^I}^I \right) \,,
\end{equation}
where 
$\lambda_1^I , \dots , \lambda_{N_w^I}^I$ are 
the eigenvalues of transfer matrix for the ideal region on the probe $I$
and
$\vv{u}_1^I , \dots , \vv{u}_{N_w^I}^I$
the eigenfunctions corresponding to 
$\lambda_1^I , \dots , \lambda_{N_w^I}^I$.
We note that the incoming solutions and outgoing ones have the same form
when there is no magnetic field.
Then the effective Hamiltonian on site $N_p +1$ for the probe $I$
can be written by
\begin{equation}
\widetilde{\cal H}_{N_P +1}^I = {\cal H}_{N_P +1}^I - V_0 F^I \,,
\end{equation}
with
\begin{equation}
F^I = U^I \Lambda^I ({U^I})^{-1} \,,
\end{equation}
where $V_0$ is the hopping term in the ideal probe.
$\widetilde{\cal H}_{N_p +1}$ can be obtained by
exchanging ${\cal H}_{i}^{I}$ for $\widetilde{\cal H}_{i}^{I}$
in Eq.(\ref{eq:Hprobe})
and setting $i=N_p +1$.

We define corresponding Green function as
\begin{eqnarray}
\hat{G} \equiv \frac{1}{E-\widetilde{\cal H}}
=
\left(
\begin{array}{ccccc}
\hat{G}_{N_p +1,N_p +1} & \hat{G}_{N_p +1,N_p} & \cdots & \hat{G}_{N_p +1,1} & 
\hat{G}_{N_p +1,C}\\
\hat{G}_{N_p,N_p +1} &\hat{G}_{N_p,N_p} & \cdots & \hat{G}_{N_p,1} & 
\hat{G}_{N_p,C}\\
\vdots&\vdots&\ddots&\vdots&\vdots\\
\hat{G}_{1,N_p +1} &\hat{G}_{1,N_p} & \cdots & \hat{G}_{1,1} & \hat{G}_{1,C}\\
\hat{G}_{C,N_p +1} &\hat{G}_{C,N_p} & \cdots & \hat{G}_{C,1} & \hat{G}_{C,C}\\
\end{array}
\right) \,.
\end{eqnarray}
In principle, we can obtain Green function by 
the direct inversion of the matrix $E-\widetilde{\cal H}$.
However, the number of numerical operations required
for the inversion of the matrix 
increases as $N^3$ with $N$ the size of the matrix.
We therefore need to reduce the size of matrix to inverse 
as small as possible.

Since we put the effective Hamiltonian $\widetilde{\cal H}_{N_p +1}$ 
on the upper-left corner of the full Hamiltonian 
(Eq.(\ref{eq:effectH})),
the sub-matrix $\hat{G}_{N_p +1,N_p+1}$ contains all information of the 
transport.
This sub-matrix can be calculated as
\begin{equation}
\hat{G}_{N_p +1,N_p +1} = \frac{1}{E-\widetilde{\cal H}_{N_p +1} - \Gamma_{N_p 
+1}} 
\,,
\end{equation}
where
\begin{equation}
\Gamma_{i+1} = V_{i+1,i} \, \frac{1}{E-{\cal H}_i -\Gamma_i} \,
V_{i+1,i}^{\dagger} \hspace{1cm} ({\rm for} \ i=1,2,\dots,N_p) \,,
\end{equation}
with
\begin{equation}
\Gamma_{1} = V'_{1,C} \, \frac{1}{E-{\cal H}^C} \, {V'}_{1,C}^{\dagger} \,.
\end{equation}
The inverse matrix of $E-{\cal H}^C$ can be calculated 
recursively\cite{Ando91} and
each row of the matrix $V'_{1,C}$ contains only one non-zero element 
corresponding to the matrix element between neighbouring slices,
which greatly simplifies the calculation of $\Gamma_1$.
In the calculation of 
$\Gamma_{i+1}$ for $i=1,2,\dots,N_p$ and $\hat{G}_{N_p +1,N_p+1}$,
we need to inverse directly the matrix whose size is
$(N_{w}^{D}+N_{w}^{R}+N_{w}^{L}) \times (N_{w}^{D}+N_{w}^{R}+N_{w}^{L})$.
This is much smaller than the original size of $E-\widetilde{\cal H}$.

$\hat{G}_{N_p +1,N_p +1}$ consists of 9 parts.
\begin{eqnarray}
\hat{G}_{N_p +1,N_p +1}
=
\left(
\begin{array}{ccc}
\hat{G}^{DD} & \hat{G}^{DR} & \hat{G}^{DL} \\
\hat{G}^{RD} & \hat{G}^{RR} & \hat{G}^{RL} \\
\hat{G}^{LD} & \hat{G}^{LR} & \hat{G}^{LL} \\
\end{array}
\right)
\end{eqnarray}
where $\hat{G}^{IJ}$ ($I,J=D,R$ and $L$) is the Green function
from the probe $J$ to $I$.
The scattering matrix can be calculated easily from the Green function
\cite{Ando91}.

\bibliography{T-shape}

\begin{thebibliography}{28}
\expandafter\ifx\csname natexlab\endcsname\relax\def\natexlab#1{#1}\fi
\expandafter\ifx\csname bibnamefont\endcsname\relax
  \def\bibnamefont#1{#1}\fi
\expandafter\ifx\csname bibfnamefont\endcsname\relax
  \def\bibfnamefont#1{#1}\fi
\expandafter\ifx\csname citenamefont\endcsname\relax
  \def\citenamefont#1{#1}\fi
\expandafter\ifx\csname url\endcsname\relax
  \def\url#1{\texttt{#1}}\fi
\expandafter\ifx\csname urlprefix\endcsname\relax\def\urlprefix{URL }\fi
\providecommand{\bibinfo}[2]{#2}
\providecommand{\eprint}[2][]{\url{#2}}

\bibitem[{\citenamefont{{S.A. Wolf {\it et al.}}}(2001)}]{Wolf}
\bibinfo{author}{\bibnamefont{{S.A. Wolf {\it et al.}}}},
  \bibinfo{journal}{Science} \textbf{\bibinfo{volume}{294}},
  \bibinfo{pages}{1488} (\bibinfo{year}{2001}).

\bibitem[{\citenamefont{{I. Zutic, J. Fabian and S.D. Sarma}}(2004)}]{Zutic}
\bibinfo{author}{\bibnamefont{{I. Zutic, J. Fabian and S.D. Sarma}}},
  \bibinfo{journal}{Rev. Mod. Phys.} \textbf{\bibinfo{volume}{76}},
  \bibinfo{pages}{323} (\bibinfo{year}{2004}).

\bibitem[{\citenamefont{{G. Dresselhaus}}(1955)}]{Dresselhaus}
\bibinfo{author}{\bibnamefont{{G. Dresselhaus}}}, \bibinfo{journal}{Phys. Rev.}
  \textbf{\bibinfo{volume}{100}}, \bibinfo{pages}{580} (\bibinfo{year}{1955}).

\bibitem[{\citenamefont{{E.I. Rashba}}(1960)}]{Rashba60}
\bibinfo{author}{\bibnamefont{{E.I. Rashba}}}, \bibinfo{journal}{Sov. Phys.
  Solid State} \textbf{\bibinfo{volume}{2}}, \bibinfo{pages}{1109}
  (\bibinfo{year}{1960}).

\bibitem[{\citenamefont{{Y.A. Bychkov and E.I. Rashba}}(1984)}]{Rashba84}
\bibinfo{author}{\bibnamefont{{Y.A. Bychkov and E.I. Rashba}}},
  \bibinfo{journal}{J. Phys. C} \textbf{\bibinfo{volume}{17}},
  \bibinfo{pages}{6039} (\bibinfo{year}{1984}).

\bibitem[{\citenamefont{{G. Lommer, F. Malcher and U.
  R\"{o}ssler}}(1988)}]{Lommer}
\bibinfo{author}{\bibnamefont{{G. Lommer, F. Malcher and U. R\"{o}ssler}}},
  \bibinfo{journal}{Phys. Rev. Lett.} \textbf{\bibinfo{volume}{60}},
  \bibinfo{pages}{728} (\bibinfo{year}{1988}).

\bibitem[{\citenamefont{{J. Nitta, T. Akazaki, H. Takayanagi and T.
  Enoki}}(1997)}]{Nitta}
\bibinfo{author}{\bibnamefont{{J. Nitta, T. Akazaki, H. Takayanagi and T.
  Enoki}}}, \bibinfo{journal}{Phys. Rev. Lett.} \textbf{\bibinfo{volume}{78}},
  \bibinfo{pages}{1335} (\bibinfo{year}{1997}).

\bibitem[{\citenamefont{{G. Engels, J. Lange, Th. Sch\"{a}pers and H.
  L\"{u}th}}(1997)}]{Engels}
\bibinfo{author}{\bibnamefont{{G. Engels, J. Lange, Th. Sch\"{a}pers and H.
  L\"{u}th}}}, \bibinfo{journal}{Phys. Rev. B} \textbf{\bibinfo{volume}{55}},
  \bibinfo{pages}{R1958} (\bibinfo{year}{1997}).

\bibitem[{\citenamefont{{S. Datta and B. Das}}(1990)}]{Datta}
\bibinfo{author}{\bibnamefont{{S. Datta and B. Das}}}, \bibinfo{journal}{Appl.
  Phys. Lett.} \textbf{\bibinfo{volume}{56}}, \bibinfo{pages}{665}
  (\bibinfo{year}{1990}).

\bibitem[{\citenamefont{{G. Schmidt, D. Ferrand, L.W. Molenkamp, A.T. Filip and
  B.J. van Wees}}(2000)}]{Schmidt}
\bibinfo{author}{\bibnamefont{{G. Schmidt, D. Ferrand, L.W. Molenkamp, A.T.
  Filip and B.J. van Wees}}}, \bibinfo{journal}{Phys. Rev. B}
  \textbf{\bibinfo{volume}{62}}, \bibinfo{pages}{R4790} (\bibinfo{year}{2000}).

\bibitem[{\citenamefont{{L.D. Landau and E.M. Lifshitz}}()}]{Landaulifshits}
\bibinfo{author}{\bibnamefont{{L.D. Landau and E.M. Lifshitz}}}, \eprint{{{\it
  Quantum mechanics: non-relativistic theory}, 3d ed., Pergamon Press, New
  York, (1991).}}

\bibitem[{\citenamefont{{A.A. Kiselev and K.W. Kim}}(2001)}]{Kiselev}
\bibinfo{author}{\bibnamefont{{A.A. Kiselev and K.W. Kim}}},
  \bibinfo{journal}{Appl. Phys. Lett.} \textbf{\bibinfo{volume}{78}},
  \bibinfo{pages}{775} (\bibinfo{year}{2001}).

\bibitem[{\citenamefont{{A.A. Kiselev and K.W. Kim}}(2003)}]{Kiselev-ring}
\bibinfo{author}{\bibnamefont{{A.A. Kiselev and K.W. Kim}}},
  \bibinfo{journal}{J. Appl. Phys.} \textbf{\bibinfo{volume}{94}},
  \bibinfo{pages}{4001} (\bibinfo{year}{2003}).

\bibitem[{\citenamefont{{M. Governale and U. Z\"{u}licke}}(2002)}]{Governale}
\bibinfo{author}{\bibnamefont{{M. Governale and U. Z\"{u}licke}}},
  \bibinfo{journal}{Phys. Rev. B} \textbf{\bibinfo{volume}{66}},
  \bibinfo{pages}{073311} (\bibinfo{year}{2002}).

\bibitem[{\citenamefont{{C.W.J. Beenakker}}(1997)}]{Beenakker}
\bibinfo{author}{\bibnamefont{{C.W.J. Beenakker}}}, \bibinfo{journal}{Rev. Mod.
  Phys.} \textbf{\bibinfo{volume}{69}}, \bibinfo{pages}{731}
  (\bibinfo{year}{1997}).

\bibitem[{\citenamefont{{A.G. Mal'shukov and K.A. Chao}}(2000)}]{Malshukov}
\bibinfo{author}{\bibnamefont{{A.G. Mal'shukov and K.A. Chao}}},
  \bibinfo{journal}{Phys. Rev. B} \textbf{\bibinfo{volume}{61}},
  \bibinfo{pages}{R2413} (\bibinfo{year}{2000}).

\bibitem[{\citenamefont{{A.A. Kiselev and K.W. Kim}}(2000)}]{Kiselev-wire}
\bibinfo{author}{\bibnamefont{{A.A. Kiselev and K.W. Kim}}},
  \bibinfo{journal}{Phys. Rev. B} \textbf{\bibinfo{volume}{61}},
  \bibinfo{pages}{13115} (\bibinfo{year}{2000}).

\bibitem[{\citenamefont{{B.K. Nikoli\'{c} and S. Souma}}(2005)}]{Nikolic-DP}
\bibinfo{author}{\bibnamefont{{B.K. Nikoli\'{c} and S. Souma}}},
  \bibinfo{journal}{Phys. Rev. B} \textbf{\bibinfo{volume}{71}},
  \bibinfo{pages}{195328} (\bibinfo{year}{2005}).

\bibitem[{\citenamefont{{S. Pramanik, S. Bandyopadhyay and M.
  Cahay}}()}]{Pramanik}
\bibinfo{author}{\bibnamefont{{S. Pramanik, S. Bandyopadhyay and M. Cahay}}},
  \eprint{cond-mat/0403021}.

\bibitem[{\citenamefont{{T. Ando}}(1989)}]{Ando89}
\bibinfo{author}{\bibnamefont{{T. Ando}}}, \bibinfo{journal}{Phys. Rev. B}
  \textbf{\bibinfo{volume}{40}}, \bibinfo{pages}{5325} (\bibinfo{year}{1989}).

\bibitem[{\citenamefont{{R. Landauer}}(1957)}]{Landauer}
\bibinfo{author}{\bibnamefont{{R. Landauer}}}, \bibinfo{journal}{IBM J. Res.
  Dev.} \textbf{\bibinfo{volume}{1}}, \bibinfo{pages}{223}
  (\bibinfo{year}{1957}).

\bibitem[{\citenamefont{{T. Ando}}(1991)}]{Ando91}
\bibinfo{author}{\bibnamefont{{T. Ando}}}, \bibinfo{journal}{Phys. Rev. B}
  \textbf{\bibinfo{volume}{44}}, \bibinfo{pages}{8017} (\bibinfo{year}{1991}).

\bibitem[{\citenamefont{{R. Karplus and J. Luttinger}}(1954)}]{Karplus}
\bibinfo{author}{\bibnamefont{{R. Karplus and J. Luttinger}}},
  \bibinfo{journal}{Phys. Rev.} \textbf{\bibinfo{volume}{95}},
  \bibinfo{pages}{1154} (\bibinfo{year}{1954}).

\bibitem[{\citenamefont{{J. Li, L. Hu and S.-Q. Shen}}()}]{li05}
\bibinfo{author}{\bibnamefont{{J. Li, L. Hu and S.-Q. Shen}}},
  \eprint{cond-mat/0502102}.

\bibitem[{\citenamefont{{B.K. Nikoli\'{c}, L.P. Z$\hat{\rm a}$rbo and S.
  Welack}}()}]{Nikolic}
\bibinfo{author}{\bibnamefont{{B.K. Nikoli\'{c}, L.P. Z$\hat{\rm a}$rbo and S.
  Welack}}}, \eprint{cond-mat/0503415, to appear in Phys. Rev. B.}

\bibitem[{\citenamefont{{C.W.J. Beenakker and H. van
  Houten}}(1989)}]{beenakker89}
\bibinfo{author}{\bibnamefont{{C.W.J. Beenakker and H. van Houten}}},
  \bibinfo{journal}{Phys. Rev. Lett.} \textbf{\bibinfo{volume}{63}},
  \bibinfo{pages}{1857} (\bibinfo{year}{1989}).

\bibitem[{\citenamefont{{S. Murakami, N. Nagaosa and S.-C.
  Zhang}}(2003)}]{Murakami}
\bibinfo{author}{\bibnamefont{{S. Murakami, N. Nagaosa and S.-C. Zhang}}},
  \bibinfo{journal}{Science} \textbf{\bibinfo{volume}{301}},
  \bibinfo{pages}{1348} (\bibinfo{year}{2003}).

\bibitem[{\citenamefont{{C.-M. Hu, C. Zehnder, Ch. Heyn and D.
  Heitmann}}(2003)}]{Hu}
\bibinfo{author}{\bibnamefont{{C.-M. Hu, C. Zehnder, Ch. Heyn and D.
  Heitmann}}}, \bibinfo{journal}{Phys. Rev. B} \textbf{\bibinfo{volume}{67}},
  \bibinfo{pages}{201302(R)} (\bibinfo{year}{2003}).

\end{thebibliography}

\end{document}